\documentclass[12pt]{article}
\usepackage{graphicx}
\begin{document}
 \noindent {\footnotesize\it Astronomy Letters, 2010 Vol. 36, No. 1,  pp. 27-43}

 \noindent
 \begin{tabular}{llllllllllllllllllllllllllllllllllllllllllllll}
 & & & & & & & & & & & & & & & & & & & & & & & & & & & & & & & & & & & & &  \\\hline\hline
 \end{tabular}

 \vskip 1.5cm
 \centerline {\large\bf Analysis of Peculiarities of the Stellar Velocity Field }
 \centerline {\large\bf in the Solar Neighborhood}
 \bigskip
 \centerline {V.V. Bobylev$^1$, A.T. Bajkova$^1$, and A. A. Myll\"{a}ri$^2$}
 \bigskip
 \centerline {\small\it
 $^1$Pulkovo Astronomical Observatory, Russian Academy of Sciences,
 St-Petersburg}
 \centerline {\small\it $^2$Turku University, Turku, Finland}
 \bigskip

{\bf Abstract--}Based on a new version of the Hipparcos catalogue
and an updated Geneva-Copenhagen survey of F and G dwarfs, we
analyze the space velocity field of $\approx$17000 single stars in
the solar neighborhood. The main known clumps, streams, and
branches (Pleiades, Hyades, Sirius, Coma Berenices, Hercules, Wolf
630-$\alpha$Ceti, and Arcturus) have been identified using various
approaches. The evolution of the space velocity field for F and G
dwarfs has been traced as a function of the stellar age. We have
managed to confirm the existence of the recently discovered KFR08
stream. We have found 19 Hipparcos stars, candidates for
membership in the KFR08 stream, and obtained an isochrone age
estimate for the stream, 13 Gyr. The mean stellar ages of the Wolf
630-$\alpha$Ceti and Hercules streams are shown to be comparable,
4--6 Gyr. No significant differences in the metallicities of stars
belonging to these streams have been found. This is an argument
for the hypothesis that these streams owe their origin to a common
mechanism.

\medskip

\noindent DOI: 10.1134/S1063773710010044

\section*{INTRODUCTION}

Studying the stellar velocity field in the solar neighborhood is
of great importance in understanding the kinematics and evolution
of various structural components in the Galaxy. At present, it is
well known that the stellar space velocity distribution has a
complex small-scale structure. This may be attributable to various
dynamical factors (the influence of a spiral density wave, the
Galactic bar, etc.).

The stellar velocity field in the solar neighborhood was analyzed
by Chereul et al. (1998),Dehnen (1998), Asiain et al. (1999),
Skuljan et al. (1999), and Torra et al. (2000) using Hipparcos
(ESA 1997) data. The space velocities of K and M giants were
studied by Famaey et al. (2005) using data from the Hipparcos and
Tycho-2 (Hog et al., 2000) catalogues in combination with the
radial velocities measured by the CORAVEL spectrovelocimeter.
Based on data from the first version of the Geneva-Copenhagen
survey (Nordstr\"{o}m et al., 2004), Bobylev and Bajkova (2007a)
analyzed the space velocities of F and G dwarfs as a function of
the stellar age. Antoja et al. (2008) studied an extensive sample
of stars of various spectral types, from O to M, using the stellar
ages and space velocities.

The theory of stellar streams has long been used to explain the
nature of the observed inhomogeneity of the stellar velocity
field. Therefore, the names to the peaks were given by association
with open star clusters (OSCs), such as the Pleiades (with an age
of 70--125 Myr; Soderblom et al. 1993), the Sirius-Ursa Majoris
cluster (500 Myr; King et al. 2003), or the Hyades (650 Myr;
Castellani et al. 2001).

The theory of stellar streams suggests a common origin of the
stars in a specific stream (Eggen 1996). The clumpy structure of
the observed velocity field in the solar neighborhood is explained
by a superposition of stars belonging to different streams.

As numerical simulations of the dynamical evolution of such OSCs
as the Hyades, the Pleiades, and Coma Berenices show (Chumak et
al. 2005; Chumak and Rastorguev 2006a, 2006b), stellar tails
elongated along the Galactic orbit of the cluster appear during
their evolution. However, in a time $\approx$2 Gyr, the OSC
remnants existing in the form of tails must completely disperse
and mix with the stellar background (K\"{u}pper et al. 2008).

The theory of stellar streams runs into great difficulties in
explaining the existence of peaks or clumps in velocity space
containing old (older than 2--4 Gyr) stars. Analysis of the
stellar metallicities performed by Taylor (2000) for nine old
streams (Hercules, Wolf 630, 61 Cyg, Arcturus, HR 1614, and
others) composed according to Eggen's lists showed such a large
spread in metallicity that a common origin of the stars in each of
the streams is out of the question. With regard to HR 1614, there
is still the opinion based on the chemical homogeneity of the
stars that this is an OSC remnant with an age of about 2 Gyr (De
Silva et al. 2007).

In recent years, nonaxisymmetric models of the Galaxy (a spiral
structure, a bar, a triaxial halo) have been invoked to account
for peculiarities in the distribution of stellar velocities in the
solar neighborhood. For example, the Galactic spiral structure
gives rise to clumpiness in the observed velocity field (De Simone
et al. 2004; Quillen and Minchev 2005). The bar at the Galactic
center (Dehnen 1999, 2000; Fux 2001; Chakrabarty 2007) leads to a
bimodal distribution of the observed $UV$ velocities.

At present, clumps of a completely different nature to which the
Sirius, Hercules, and Arcturus streams belong are distinguished.

In the opinion of Klement et al. (2008), the Sirius stream
contains not only stars formed simultaneously and evolving as an
OSC but also a sizeable fraction of field stars that fell into
this region through the impact of a spiral density wave.

Numerical simulations have shown that the existence of the
Hercules stream $(V\approx-50$ km s$^{-1})$ can be explained by
the fact that its stars have resonant orbits induced by the
Galactic bar (Dehnen 1999, 2000; Fux 2001). In this case, the Sun
must be located near the outer Lindblad resonance. A detailed
analysis performed by Bensby et al. (2007) using high-resolution
spectra of nearby F and G dwarfs showed this stream to contain
stars of various ages, metallicities, and elemental abundances.
Bensby et al. (2007) concluded that the influence of a bar-type
dynamical factor is the most acceptable explanation for the
existence of the Hercules stream.

Several authors (Navarro et al. 2004; Helmi et al. 2006; Arifyanto
and Fuchs 2006) concluded that the Arcturus stream $(V\approx-100$
km s$^{-1})$ is the old ($\approx$15 Gyr) debris of a dwarf galaxy
captured by the Galaxy and disrupted by its tidal effect. Data on
the kinematics and metallicities of the stars being analyzed
served as arguments for this conclusion.

Analysis of the RAVE DR1 experimental data (Steinmetz et al. 2006)
revealed a hitherto unknown stream (Klement et al. 2008) with an
age of $\approx13$ Gyr in the region of ``rapidly flying'' stars
$(V\approx-160$ km s$^{-1})$ whose origin has not yet been
established.

The goal of this paper is to analyze peculiarities of the stellar
velocity field in the solar neighborhood based on a new version of
the Hipparcos catalogue, the OSACA and PCRV catalogs of radial
velocities, and an updated Geneva-Copenhagen survey of F and G
dwarfs, which provide the currently most accurate data on the
individual distances, space velocities, and ages of stars.

\section*{THE COORDINATE SYSTEM}

In this paper, we use a rectangular Galactic coordinate system
with the axes directed away from the observer toward the Galactic
center
 $(l=0^\circ,$ $b=0^\circ,$ the $X$ axis), along the Galactic rotation
 $(l=90^\circ,$ $b=0^\circ,$ the $Y$ axis), and toward the North Galactic Pole
 $(b=90^\circ,$ the $Z$ axis). The corresponding space velocity components
of the object $U,V,$ and $W$ are also directed along the $X,Y,$
and $Z$ axes.

\section*{THE DATA}

We use stars from the Hipparcos catalog (ESA 1997). We took the
proper-motion components and parallaxes from an updated version of
the Hipparcos catalog (van Leeuwen 2007), the stellar radial
velocities from the OSACA compilation catalog of radial velocities
(Bobylev et al. 2006) and the Pulkovo Compilation of Radial
Velocities (Gontcharov 2006); improved age estimates and
metallicity indices [Fe/H] for F and G dwarfs were taken from an
updated Geneva-Copenhagen survey (Holmberg et al. 2007, 2008).

As a result, we have data of various quality on 34359 stars of
various spectral types. Among them, 16737 stars are single ones
with the most reliable distance estimates, i.e., $e_\pi/\pi<0.1$
for them. We chose the constraint on the parallax errors from the
considerations of selecting a sufficiently large number of stars
at the minimal effect of Lutz and Kelker (1973). These stars
constitute our main working sample that we designate as ``all''
(Figs. 1, 2, 4, 5). The stellar UV-velocity distribution for this
sample is presented in Fig.~1a.

For the selected stars, we, nevertheless, made a statistical
estimate of the $U$ and $V$ velocity biases caused by the
measurement errors of the stellar parallaxes. For this purpose, we
used the method of Monte Carlo simulations. We generated 1000
random realizations of parallax errors for each star that
satisfied a normal law. Figures 1b and 1c present the derived
histograms separately for the U and V velocities, respectively.
The number of stars whose velocity bias lies in a certain bin
along the horizontal axis is indicated along the vertical axis. As
we see from the histograms, the statistical U and V velocity
biases caused by the parallax errors are generally insignificant;
for 70\% of the stars, they lie in the interval $[-0.05,0.05]$ km
s$^{-1}$. The maximum bias (given the asymmetry of the derived
distributions) does not exceed 0.5 km s$^{-1}$. This value is
approximately a factor of 2--3 lower than the statistical
uncertainty caused by the measurement errors of the proper motions
and radial velocities (Skuljan et al. 1999).

The stellar velocities were corrected for the differential
rotation of the Galaxy. The Galactic differential rotation effect
is known to manifest itself in its influence on the $U$ velocity
via the gradient $dU/dY = -\Omega_0,$ then $\Delta U = (dU/dY )Y =
-\Omega_0 Y ,$ where $\Omega_0 = B-A\approx-30$ km s$^{-1}$
kpc$^{-1}$. This means that for a typical error in the stellar
space velocities of $\varepsilon = 1$ km s$^{-1}$, this effect may
be disregarded only for the stars within $d < \varepsilon/\Omega_0
= 33$ pc. Since the stars used also have greater distances, the
differential rotation of the Galaxy should be taken into account.

The Galactic rotation parameters (the Oort constants A and B) have
been repeatedly determined by various authors (Zabolotskikh et al.
2002; Olling and Dehnen 2003; Bobylev 2004); they are known with
an error $\sigma\approx 1$ km s$^{-1}$ kpc$^{-1}$. This means that
for a typical error in the stellar space velocities of
$\varepsilon = 1$ km s$^{-1}$, the influence of an uncertainty in
determining $\Omega_0$ is significant for the stars located at
distances $d> \varepsilon/3\sigma = 333$ pc. Fortunately, the
number of such distant stars in our ``all'' sample is small (only
two or three dozen OB stars), and their influence may be
neglected. In this paper, we use the Oort constants $A =
13.7\pm0.6$ km s$^{-1}$ kpc$^{-1}$ and $B = -12.9\pm0.4$ km
s$^{-1}$ kpc$^{-1}$ that were determined by Bobylev (2004) from an
analysis of the independent estimates obtained by various authors.

\section*{THE METHODS}

\subsection*{The Adaptive Kernel Method}

We use an adaptive kernel method to obtain an estimate of the
velocity distribution $f(U,V)$ similar to that of the probability
density distribution from the initial velocity distribution
presented in Fig. 1. In contrast to the approach of Skuljan et al.
(1999), we use a two-dimensional, radially symmetric Gaussian
kernel function expressed as
\begin{equation}
K(r,\sigma)=\frac{1}{2\pi\sigma^2}
\exp\Biggl(-{\frac{r^2}{2\sigma^2}}\Biggr),
\end{equation}
where $r^2=x^2+y^2$ and $\sigma$ is a positive bandwidth
parameter; in this case, the relation $\int K(r)dr=1$ needed to
estimate the probability density holds. Obviously, the larger the
parameter $\sigma$, the larger the bandwidth and the lower the
amplitude.

The basic idea of the adaptive kernel method is that at each point
of the map, the operation of convolution with a band of the width
specified by the parameter $\sigma$ that varies in accordance with
the data density near this point is performed. Thus, in zones with
an enhanced density, the smoothing is done by a comparatively
narrow band; the bandwidth increases with decreasing data density.

We will use the following definition of the adaptive kernel
estimator at an arbitrary point $\xi=(U,V)$ (Silverman 1986;
Skuljan et al. 1999) adapted to a Gaussian kernel function:
$$
 \hat{f}(\xi)=\frac{1}{n}\sum_{i=1}^n K\left(|\xi-\xi_i|,{h\lambda_i}\right),
$$
where $\xi_i=(U_i,V_i), \lambda_i$ is the local dimensionless
bandwidth parameter at point $\xi_i, h$ is a general smoothing
parameter, $n$ is the number of data points $\xi_i = (U_i, V_i).$
The parameter $\lambda_i$ at each point of the two-dimensional
$UV$ plane is defined as
\begin{equation}
\lambda_i=\sqrt\frac{g}{\hat{f}(\xi_i)},
\end{equation}
  where $g$ is the
geometric mean of $\hat{f}(\xi_i)$:
\begin{equation}
\ln g=\frac{1}{n}\sum_{i=1}^n \ln \hat{f}(\xi_i).
\end{equation}
Obviously, to determine $\lambda_i$ from Eqs. (2)--(3), we must
know the distribution $\hat{f}(\xi)$ which, in turn, can be
determined if all $\lambda_i$ are known. Therefore, the problem of
finding the sought-for distribution is solved iteratively. As the
first approximation, we use the distribution obtained by smoothing
the initial $UV$ map with a band of an arbitrary fixed width. The
optimal value of the parameter $h$ can be found from the condition
for the rms deviation of the estimator $\hat{f}(\xi)$ from the
true distribution $f(\xi)$ being at a minimum. In contrast to
Skuljan et al. (1999), to determine $\lambda_i$ at each iteration,
we used the values of the function $\hat{f}(\xi)$ determined not
at the specified points  $\xi_i$ but at all points of an
equidistant grid on which the smoothed $UV$ distribution is
sought. As our comparison showed, both smoothing methods yield
approximately the same results, but, at the same time, our
approach requires much less computation. The value of $h$ for all
maps was taken to be 5.0. To obtain each map, we made 20
iterations.

The sampling interval of the two-dimensional maps was chosen from
a typical uncertainty in the $U$ and $V$ velocities (Skuljan et
al. 1999). In our case, it is 2 km s$^{-1}$, since the velocity
errors for most of the stars in the solar neighborhood (about
80\%) do not exceed $\pm$1 km s$^{-1}$. The sampling interval of
the maps in our analysis of the velocity distributions for age
separated samples was taken to be $d = 2$ km s$^{-1}$. In
analyzing the ``all'' sample of stars, we chose $d = 1$ km
s$^{-1}$ from a large number of stars as an optimal one from the
standpoint of providing the necessary detail of the derived
smoothed distribution. To obtain distributions similar to the
probability density distribution, the smoothed two-dimensional
velocity distributions must be scaled by the factor $n\times s,$
where $s = d \times d$ km$^2$ s$^2.$ The map size was $256\times
256$ pixels at the square bin size $s = 2\times 2 = 4$ km$^2$
s$^2$ in the first case and $512\times 512$ pixels at $s =
1\times1 = 1$ km$^2$ s$^2$ in the second case.

\subsection*{Wavelet Analysis}

To identify statistically significant signals of the main
inhomogeneities in the distributions of $UV$ velocities, we also
use the wavelet transform technique. This is known as a powerful
tool for filtering spatially localized signals (Chui 1997;
Vityazev 2001).

The wavelet transform of a two-dimensional distribution $f(U,V)$
consists in its decomposition into analyzing wavelets $\psi(U/a,
V/a),$ where $a$ is the scale parameter that allows a wavelet of a
particular scale to be selected from the entire family of wavelets
characterized by the same shape $\psi$. The wavelet transform
$w(\xi,\eta)$ is defined as a correlation function, so that we
have one real value of the following integral at any given point
$(\xi,\eta)$ in the $UV$ plane:
$$
w(\xi,\eta)=\int_{-\infty}^\infty \int_{-\infty}^\infty f(U,V)
\psi\Biggl(\frac{(U-\xi)}{a},\frac{(V-\eta)}{a}\Biggr) dU dV,
$$
which is called the wavelet coefficient at $(\xi,\eta)$.
Obviously, in our case of finite discrete maps, their number is
finite and equal to the number of square bins on the map.

As the analyzing wavelet, we use a standard wavelet called a
Mexican hat (MHAT). A two-dimensional MHAT wavelet is given by
\begin{equation}
\psi(r/a)=\Biggl(2-\frac{r^2}{a^2}\Biggr)e^{-r^2/2a^2},
\end{equation}
 where $r^2=U^2+V^2.$
 Wavelet (4) is obtained by
doubly differentiating the Gaussian function. The parameter $a$
that specifies the spatial scale (width) of the wavelet $\psi$ is
analogous to the parameter $\sigma$ in Eq. (1). The main property
of the wavelet $\psi$ is that its integral over $U$ and $V$ is
equal to zero, which allows any inhomogeneities to be detected in
the investigated distribution. If the distribution being analyzed
is inhomogeneous, then all coefficients of the wavelet transform
will be zero.

For our wavelet analysis of various samples in the planes of
$UV,VW,UW$ velocities and in the $(V,\sqrt{ U^2 + 2V^2} )$ plane,
we chose the scale parameter a to be 8.37 km s$^{-1}$. The value
of this parameter allowed us to reliably identify the most
significant structural features of the velocity distribution that
are the subject of our investigation. Note that for our analysis
of the velocities in the $(V,\sqrt{ U^2 + 2V^2} )$ plane, the map
size was $1024\times1024$ pixels, with the square bin size being
 $s = 1\times 1 = 1$ km$^2$ s$^2.$

\section*{RESULTS}

Figure 2a presents the $UV$-velocity distribution for the selected
16737 single stars (the ``all'' sample) obtained by the adaptive
kernel method applied to the initial velocity distribution shown
in Fig. 1. The contour lines are drawn with a uniform step equal
to 2\% of the distribution peak.

The classical Pleiades,
 $(U,V)=(-14,-23)$ kms$^{-1},$ Hyades,
 $(U,V)=(-43,-20)$ km s$^{-1},$ Sirius,
 $(U,V)=(-8,  2)$ km s$^{-1},$ and Coma Berenices,
 $(U,V)=(-11,-8)$ km s$^{-1},$ streams as well as the Hercules,
 $(U,V)=(-31,-49)$ km s$^{-1},$ stream are clearly
distinguished in Fig. 2a. In addition, there is a blurred clump
elongated along the U axis in a wide region
 $(U,V) \approx (37,-22)$ km s$^{-1}$. In the opinion of Antoja et al. (2008), the Wolf 630 peak
 $(U,V) = (25,-33)$ km s$^{-1}$ (Eggen 1996) and the nameless peak
 $(U,V) = (50,-25)$ km s$^{-1}$ (Dehnen 1998) are associated with this new clump.
Francis and Anderson (2008) designated this clump as the
$\alpha$Ceti stream; the UV coordinates of the star $\alpha$Ceti,
 $(U,V) = (25,-23)$ km s$^{-1}$, are also far from the characteristic clump center, as for
Wolf 630. As a compromise, we suggest calling this structure the
Wolf 630-$\alpha$Ceti stream or branch.

Figure 2b presents the sections of map 2a perpendicular to the
$(U,V)$ plane that pass through the main peaks and that make
$+16^\circ$ with the $U$ axis if measured clockwise (this axis is
designated in the figure as $U);$ the distribution density in
units of $7\times10^{-4}$ is along the vertical axis. The
orientation of the sections coincides with the direction of the
``branches'' detected on the smoothed maps (see also Skuljan et
al. 1999; Antoja et al. 2008).

As we see from Fig. 2, the Hyades peak dominates in amplitude,
although the Pleiades peak is integrally more powerful, as can be
seen from the wavelet distribution for the ``all'' sample shown in
Fig. 4.

Figure 3 present the $UV$-velocity distributions for eight samples
(t1--t8) of F and G dwarfs as a function of the stellar age, which
allow the evolution of the main peaks and clumps to be traced. We
used a total of 6079 single stars with distance and age errors
$e_\pi/\pi<0.2$ and $e_\pi/\pi<0.3,$ respectively. The mean ages
$\tau$ of samples t1--t8 are 1.2, 1.7, 2.2, 2.7, 3.4, 4.9, 7.2,
and 11.2 Gyr, respectively. The numbers of stars in samples t1--t8
are 509, 1105, 1184, 823, 803, 558, 586, and 511, respectively.
The step of the contour lines in Fig. 3 is 6.7\% of the peak
value.

As we see from Fig. 3, the ratio of the amplitudes of the main
peaks changes with age. For example, for the samples of
comparatively young stars (t1,t2,t3), the Hyades peak is dominant;
the Pleiades peak is gradually enhanced with stellar age (t4,t5)
and is already dominant for sample t6. The Hyades and Pleiades
peaks form an elongated structure in the shape of a ``branch''
whose orientation remains unchanged. Such structures in the
$UV$-velocity distribution for a large number of Hipparcos stars
were first described by Skuljan et al. (1999).

Numerical simulations of the disk heating by stochastic spiral
waves performed by De Simone et al. (2004) showed that the
stratification of the UV distribution into ``branches'' and peaks
could be explained by irregularities in the Galactic potential
rather than by irregularities in the star formation rate. As was
shown by Fux (2001), the presence of a bar at the Galactic center
gives rise to branches. It is currently believed that the
formation of the Hercules branch is related precisely to the
influence of a bar.

Figure 4 presents the wavelet maps of $UV,UW,$ and $VW$ velocities
for the ``all'' sample. The contour lines are given on a
logarithmic scale: 1, 2, 4, 8, 16, 32, 64, 90, and 99\% of the
peak value. Note that only the positive contours that describe the
clump regions are shown on the maps. Since the negative values of
the wavelet distributions describe the regions of a sparse
distribution of stars, they are of no interest to us and are not
shown in the figures. Such clumps as HR 1614
 $(U,V)=(15,-60)$ km s$^{-1}$ and no. 13
 $(U,V)=(50, 0)$ km s$^{-1}$ are marked in Fig. 4
according to the list by Dehnen (1998). In addition, clumps no. 8
 $(U,V)=(-40,-50)$ km s$^{-1}$, no. 9
 $(U,V)=(-25,-50)$ km s$^{-1}$, and no. 12
 $(U,V)=(-70,-50)$ km s$^{-1}$ fall into the Hercules stream,
while clump no. 14
 $(U,V)=(50,-25)$ km s$^{-1}$ falls into the Wolf630-$\alpha$Ceti stream.
As a result, out of the 14 clumps marked in Dehnen (1998), we
cannot confirm the presence of isolated clump no. 11
 $(U,V)=(-70,-10)$ km s$^{-1}$ in the region of ``high velocity''
stars. According to Navarro et al. (2004), the Arcturus stream is
located in the fairly narrow interval $-150$ km s$^{-1}<V<-100$ km
s$^{_1}$ and in the considerably wider interval
 $-150$ km s$^{-1}<U<150$ km s$^{-1}$; thus, the region marked in Fig. 4 fits into
these limits.

Figure 4 indicates features W1 and W2 for the Wolf
630-$\alpha$Ceti branch and features H1 and H2 for the Hercules
branch. According to these data, we selected the stars belonging
to these features and calculated their mean ages and
metallicities, which are given in Table 1. For the selection of
stars, we used our probabilistic approach described in detail in
Bobylev and Bajkova (2007b). Note that our samples were comparable
in the number of stars --- 525 and 625 stars are contained in the
Wolf 630-$\alpha$Ceti and Hercules branches, respectively. To
calculate the means and dispersions listed in Table 1, we used
only the stars with available age and metallicity estimates, in
fact, these are F and G dwarfs; the constraints $e_\pi/\pi<0.1$
and $e_\pi/\pi<0.3$ were used.

The last columns in Table 1 give parameters of the $1\sigma$
ellipses: the semimajor and semiminor axes $a_i$ and $b_i$ as well
as the angle $\beta_i$ between the vertical and semimajor axes
(measured from the vertical axis clockwise). The selection of
stars with these parameters was made within the boundaries of the
$3\sigma$ ellipses.

Note that no significant concentrations of stars are observed in
the $W-U$ and $W-V$ planes outside the central ``ellipse''.

Next, we apply a technique proposed by Arifyanto and Fuchs (2006)
that consists in identifying velocity field inhomogeneities in the
plane of $V,\sqrt{U^2 + 2V^2}$ coordinates. It allows low-power
streams to be reliably identified in the range of high space
velocities.

Figure 5 shows the wavelet distributions for the ``all'' sample in
the $(V, \sqrt{ U^2 + 2V^2})$ plane. The contour lines are given
on a logarithmic scale: 0.05, 0.1, 0.2, 0.4, 0.8, 1.6, 3.2, . . .,
50, and 99\% of the peak value. In Figs. 5--8, we give the stellar
velocities relative to the local standard of rest (LSR) whose
coordinates are $(U,V,W)=(10.0,5.2,7.2)$ km s$^{-1}$ (Dehnen and
Binney 1998); the cited coordinates of the clumps are also given
relative to the LSR. Figure 5 marks the AF06 stream with
coordinates $(-80,130)$ km s$^{-1}$ (Arifyanto and Fuchs 2006),
the Arcturus stream with coordinates $(-125,185)$ km s$^{-1}$
(Arifyanto and Fuchs 2006), and the KFR08 stream with coordinates
$(-160,225)$ km s$^{-1}$ (Klement et al. 2008). On this diagram,
the Wolf 630-$\alpha$Ceti stream mergers with the Hyades-Pleiades
branch.

Figure 6 presents the wavelet maps in $V,\sqrt{U^2 + 2V^2}$
coordinates for samples of F and G dwarfs as a function of the
stellar age; the set of levels is similar to that in Fig. 5. As we
see from the figure, a prominent clump of KFR08 stream stars is
observed for sample t8, which includes the oldest stars
considered. The central point in the KFR08 region marked on the
plot (t8) has the eighth level; all of the remaining clumps at
 $\sqrt{U^2 + 2V^2}>250$ km s$^{-1}$ have one level fewer and, hence, their
significance is considerably lower.

Still, it is interesting to note that there is a clump close to
the KFR08 region in Fig. 6 for sample t4. However, the
significance of the levels in this case is negligible,
corresponding to the presence of only one or two stars. A special
search showed that one star from sample t4, HIP 77946, for which
[Fe/H]$=-0.83$ and $\tau=2.5$ Gyr (Holmberg et al. 2007), falls
into the neighborhood of KFR08 with a radius of 30 km s$^{-1}$.

Table 2 gives parameters of the stars that are probable members of
the KFR08 stream.We selected the candidates for membership in this
stream based on the distribution of the expanded ``all'' sample
with $e_\pi/\pi<0.15$ in the plane of $(V, \sqrt{ U^2 + 2V^2})$
coordinates. As a result, 19 stars were selected from the
neighborhood of the clump center with coordinates $(-159,227)$ km
s$^{-1}$ and a neighborhood radius of 30 km s$^{-1}$.

To determine the probability that each of the selected stars
belonged to the KFR08 and Arcturus streams, we performed Monte
Carlo simulations of the distribution of stars in the plane of
 $V,\sqrt{ U^2 + 2V^2}$ coordinates by taking into account the random errors
in the stellar space velocities. We generated 3000 random
realizations for each star. In our simulations of the KFR08 and
Arcturus streams, we took the following parameters of their
distribution in the $V,\sqrt{U^2 + 2V^2}$ plane obtained by
analyzing Fig. 5: (1) the coordinates of the centers are
 $(-159,227)$ km s$^{-1}$ for KFR08 and
 $(-124,178)$ km s$^{-1}$ for Arcturus; (2) the
velocity dispersion is 5 km s$^{-1}$ for both streams. The results
of our simulations are reflected in Fig. 7 and in the last column
of Table 2, which gives the probability that a star belongs to the
KFR08 stream, p. Obviously, the probability that a star belongs to
the Arcturus stream is $1-p.$ As we see from Table 2, eleven stars
constituting the core of the KFR08 stream have probabilities
 $p\geq0.99$ and only two stars have $p=0.65.$ The positions of these two
stars (HIP 74033 and HIP 58357) are marked in Fig. 7. As we see
from the figure, their random errors are such that they have
almost equal chances of being attributed to both the KFR08 and
Arcturus streams. Therefore, it is not surprising that the star
HIP 74033 in Arifyanto and Fuchs (2006) was attributed to the
Arcturus stream.

Since we have failed to find information about the metallicities
of several stars from this sample in the literature, we calculated
the metallicity indices based on Str\"{o}mgren uvby? photometry
from the compilation by Hauck and Mermilliod (1998) using the
calibration by Schuster and Nissen (1989).

The distribution of $U,V,W$ velocities for KFR08 stream members is
shown in Fig. 8. As can be seen from this figure, the stars are
located in a narrow range of velocities V and in wider ranges of
 $U$ and $W$ than are typical of the Arcturus stream stars (Navarro et
al. 2004).

Figure 9 presents a color-absolute magnitude diagram for KFR08
stream members with the Yonsei-Yale (Yi et al. 2003) 11-, 13-, and
15-Gyr isochrones for $Z=0.007$ (Fe/H$=-0.43).$ We can see that
the stream stars fall nicely on the 13-Gyr isochrone; the
deviations are most pronounced only for two stars, HIP 87101 and
HIP 93269. Our isochrone age estimate for the stream is in good
agreement with the available age estimates for individual stars
(Table 2).


\section*{DISCUSSION}

(1) Using currently available data, we have been able to confirm
the presence of main known clumps, streams, and branches in the
stellar velocity field in the solar neighborhood and to trace the
evolution of the velocity field for F and G dwarfs as a function
of the stellar age. Note that there is a very wide range of
stellar ages in each of the classical Pleiades, Hyades, Sirius,
Coma Berenices, and Hercules streams (Fig. 3). This is in good
agreement with the results of a detailed analysis of the
metallicity distribution and age estimates for stars performed
recently by Antoja et al. (2008) and Francis and Anderson (2008).

(2) The Wolf 630-$\alpha$Ceti and Hercules streams are interesting
in that they both could be produced by a common mechanism related
to the influence of a bar at the Galactic center (Dehnen 1999,
2000; Fux 2001; Chakrabarty 2007). As can be seen from Fig. 3,
both streams begin to manifest themselves at a mean age of the
sample stars $>2$ Gyr. They are most pronounced at a mean stellar
age of $\approx$7 Gyr (sample t7). Using improved stellar age
estimates from an updated version of the Geneva-Copenhagen survey
(Holmberg et al. 2007, 2008) led to a noticeable shift of the mean
stellar age for the Hercules branch in the direction of its
decrease. For example, in Bobylev and Bajkova (2007a), where the
age estimates from the first version of the catalog (Nordstr\"{o}m
et al. 2004) were used, a similar development of the Hercules
branch was achieved at a mean age of the sample stars $\approx$8.9
Gyr.

According to the data by Taylor (2000), the mean stellar
metallicity is [Fe/H]$=-0.11\pm0.02\pm0.15$ dex (the error of the
mean and dispersion) for the Wolf 630 stream ($\approx$40 stars
selected according to Eggen's lists) and [Fe/H]$=
-0.12\pm0.04\pm0.18$ dex (the error of the mean and dispersion)
for the Hercules stream ($\approx$10 stars).

An extensive analysis of the distribution of stars in age and
metallicity in various streams performed recently by Antoja et al.
(2008) showed that the highest (compared to other branches)
stellar metallicity dispersion is characteristic of the Hercules
branch. The mean and dispersion are [Fe/H]$=-0.15\pm0.27$ dex.

This structure was shown to be distinguished increasingly clearly
in the form of a branch starting from an age of 2 Gyr. Our results
are generally in good agreement with those of Antoja et al.
(2008).

The mean stellar metallicity and age for features H1 and H2 of the
Hercules stream as well as W1 and W2 of the Wolf 630-$\alpha$Ceti
branch (Table 1) are consistent with the hypothesis of a dynamical
nature of the streams related to the influence of the Galactic
bar. This is seen most clearly for features H1 and H2. Thus, for
example, feature H1, which is closer to the local standard of
rest, is youngest. Since young field stars fall into the samples
under consideration, the mean ages of the streams are
underestimated, especially for features W1 and W2.

Note that the existence of the HR 1614 clump cannot be explained
only by the presence of a OSC remnant with an age of $\sim2$ Gyr
(De Silva et al. 2007), since this clump is traceable in the $UV$
distributions for samples of considerably older stars. Thus, for
example, it is clearly seen on the $UV$ map for stars with an age
of $\approx7$ Gyr (t7, Figs. 3 and 4), suggesting that the HR 1614
clump is an outgrowth of the Hercules branch and can be dynamical
in nature.

(3) The KFR08 stream was discovered by Klement et al. (2008) from
their analysis of the data on faint (compared to Hipparcos) stars
of the RAVE experiment. These authors identified 15 stream
candidates. Since the distances of the stars in the analyzed
sample were estimated from photometric data, they are less
reliable than the trigonometric distances of Hipparcos stars. At
the same time, Klement et al. (2008) analyzed 13440 stars from the
first version of the Geneva-Copenhagen survey (Nordstr\"{o}m et
al. 2004) and showed that the presence of about 30 stars (among
the Hipparcos stars) in the KFR08 clump might be expected in the
$V,\sqrt{U^2+2V^2}$ plane. However, no specific stars were
selected.

The number of candidates for membership in the KFR08 stream we
found is in satisfactory agreement with the expected estimates.
The results of our search based on more accurate data are of great
interest in establishing the nature of the KFR08 stream. In
contrast to the samples by Klement et al. (2008), our ``all''
sample contains not only dwarfs but also giants.

As a result, we can see the main-sequence turnoff on the
color-absolute magnitude diagram for KFR08 stream members (Fig.
9), which increases the reliability of the stream age estimate
 $(\approx13$ Gyr).

According to the available data (Table 2), the metallicity indices
for an overwhelming majority of stars lie within a fairly narrow
range, $-1<$[Fe/H]$<-0.3.$ A similar homogeneity is also observed
for the stars of the Arcturus stream (Navarro et al. 2004). This
is one of the arguments for a common nature of these two streams.
Obviously,much greater statistics is required to make the final
decision.

Note that Minchev et al. (2009) suggested an alternative
hypothesis about the nature of the AF06, Arcturus, and KFR08
streams. It is based on the assumption that the disk has not yet
relaxed and it is ``shaken'' after the disruption of the dwarf
galaxy captured by our Galaxy; therefore, waves are observed in
the plane of $UV$ velocities.

\section*{CONCLUSIONS}

Based on the most recent data, we studied the space velocity field
of $\approx$17000 stars in the solar neighborhood.We used data
from a new version of the Hipparcos catalogue (van Leeuwen 2007),
stellar radial velocities from the OSACA (Bobylev et al. 2006) and
PCRV (Gontcharov 2006) catalogs reduced to a common system, and
improved estimates of the ages and metallicity indices for F and G
dwarfs from an updated Geneva-Copenhagen survey (Holmberg et al.
2007, 2008).

We identified all of the main clumps, streams, and branches known
to date using various approaches. Among the stars with a
relatively low velocity dispersion, these are the Pleiades,
Hyades, and Coma Berenices streams or branches. Among the stars
with an intermediate velocity dispersion, these are the Hercules
and Wolf 630-$\alpha$Ceti branches. Among the stars with a high
velocity dispersion, these are the Arcturus and AF06 streams
(Arifyanto and Fuchs 2006) and the KFR08 stream (Klement et al.
2008).

Our attention was focused on the most poorly studied structures,
the Wolf 630-$\alpha$Ceti and Hercules branches, and on the KFR08
stream discovered quite recently.

The present view of the nature of the Wolf 630-$\alpha$Ceti and
Hercules streams is that they could be produced by the same
mechanism related to the influence of a bar at the Galactic
center. Indeed, these structures begin to manifest themselves as
independent branches at a mean age of the sample stars $>2$ Gyr,
which is in conflict with the hypothesis of their origin based on
the theory of stellar streams (Eggen's hypothesis). Our estimates
showed that the mean stellar ages of these structures are quite
comparable and are 4--6 Gyr. We revealed now significant
differences in the metallicities of the stars belonging to these
streams.

We found 19 Hipparcos stars belonging to the new KFR08 stream and
obtained an isochrone age estimate for the stream, 13 Gyr. The
homogeneity of the kinematics, chemical composition, and age of
the sample stars is consistent with the hypothesis that the stream
is a relic remnant of the galaxy captured and disrupted by the
tidal effect of our own Galaxy. Data from the GAIA experiment will
undoubtedly play a major role for a further study of this
structure.

 \bigskip
{\bf ACKNOWLEDGMENTS}
 \bigskip

We are grateful to the referees for helpful remarks that
contributed to a improvement of the paper. The SIMBAD search
database was very helpful in the work. This study was supported by
the Russian Foundation for Basic Research (project no.
08--02--00400) and in part by the ``Origin and Evolution of Stars
and Galaxies'' Program of the Presidium of the Russian Academy of
Sciences and the Program for State Support of Leading Scientific
Schools of Russia (NSh--6110.2008.2).

 \bigskip
{\bf REFERENCES}
 \bigskip

1. T. Antoja, F. Figueras, D. Fern\'andez, and J. Torra, Astron.
Astrophys. 490, 135 (2008).

2. M. I. Arifyanto and B. Fuchs, Astron. Astrophys. 449, 533
(2006).

3. R. Asiain, F. Figueras, J. Torra, et al., Astron. Astrophys.
341, 427 (1999).

4. T. Bensby, S. Feltzing, I. Lundst\"{o}em, et al., Astron.
Astroph. 433, 185 (2005).

5. T. Bensby, M.S. Oey, S. Feltzing, et al., Astroph. J. 655, L89
(2007).

6. V.V. Bobylev, Pis'ma Astron. Zh. 30, 185 (2004) [Astron. Lett.
30, 159 (2004)].

7. V.V. Bobylev and A.T. Bajkova, Astron. Zh. 84, 418 (2007a)
[Astron. Rep. 51, 372 (2007)].

8. V.V. Bobylev and A.T. Bajkova, Pis'ma Astron. Zh. 33, 643
(2007b) [Astron. Lett. 30, 571 (2004)].

9. V.V. Bobylev, G.A. Gontcharov, and A.T. Bajkova, Astron. Zh.
83, 821 (2006) [Astron. Rep. 50, 733 (2006)].

10. T.V. Borkova and V.A. Marsakov, Astron. Zh. 82, 453 (2005)
[Astron. Rep. 49, 405 (2005)].

11. V. Castellani, S. Degl'Innocenti, and P.Moroni, Mon. Not. R.
Astron. Soc. 320, 66 (2001).

12. D. Chakrabarty, Astron. Astrophys. 467, 145 (2007). 13. E.
Chereul, M. Cr\'ez\'e, and O. Bienaym\'e, Astron. Astrophys. 340,
384 (1998).

14. C.K. Chui, Wavelets: A Mathematical Tool for Signal Analysis
(SIAM, Philadelphia, PA, 1997).

15. Ya.O. Chumak and A.S. Rastorguev, Pis'ma Astron. Zh. 32, 117
(2006a) [Astron. Lett. 32, 157 (2006)].

16. Ya.O. Chumak and A.S. Rastorguev, Pis'ma Astron. Zh. 32, 497
(2006b) [Astron. Lett. 32, 446 (2006)].

17. Ya.O. Chumak, A.S. Rastorguev, and S.J. Aarseth, Pis'ma
Astron. Zh. 31, 342 (2005) [Astron. Lett. 31, 308 (2005)].

18. W. Dehnen, Astron. J. 115, 2384 (1998).

19. W. Dehnen, Astrophys. J. 524, L35 (1999).

20. W. Dehnen, Astron. J. 119, 800 (2000).

21. W. Dehnen and J.J. Binney, Mon. Not. R. Astron. Soc. 298, 387
(1998).

22. O.J. Eggen, Astron. J. 112, 1595 (1996).

23. B. Famaey, A. Jorissen, X. Luri, et al., Astron.Astrophys.
430, 165 (2005).

24. C. Francis and E. Anderson, astro-ph: 0812.4032 (2008). 25. R.
Fux, Astron. Astrophys. 373, 511 (2001).

26. G.A. Gontcharov, Pis'ma Astron. Zh. 32, 844 (2006) [Astron.
Lett. 32, 759 (2006)].

27. B. Hauck and M. Mermilliod, Astron. Astrophys. Suppl. Ser.
129, 431 (1998).

28. A. Helmi, J.F. Navarro, B. Nordstr\"{o}m, et al., Mon. Not. R.
Astron. Soc. 365, 1309 (2006).

29. The HIPPARCOS and Tycho Catalogues, ESA SP-1200 (1997).

30. E. Hog, C. Fabricius, V.V. Makarov, et al., Astron. Astrophys.
355, L27 (2000).

31. J. Holmberg, B. Nordstr\"{o}m, and J. Andersen, Astron.
Astrophys. 475, 519 (2007).

32. J. Holmberg, B. Nordstr\"{o}m, and J. Andersen, astroph:
0811.3982 (2008).

33. A. Ibukiyama, and N. Arimoto, Astron. Astrophys. 394, 927
(2002).

34. J.S. Jenkins, H.R.A. Jones, Y. Pavlenko, et al., Astron.
Astrophys. 485, 571 (2008).

35. J.R. King, A.R. Villarreal, D.R. Soderblom, et al., Astron. J.
125, 1980 (2003).

36. R. Klement, B. Fuchs, and H.-W. Rix, Astrophys. J. 685, 261
(2008).

37. A.H.W. K\"{u}pper,A. Macleod, and D. C.Heggie, Mon. Not. R.
Astron. Soc. 387, 1248 (2008).

38. F. van Leeuwen, Astron. Astrophys. 474, 653 (2007).

39. T.E. Lutz and D.H. Kelker, Publ. Astron.Soc.Pacific 85, 573
(1973).

40. I. Minchev, A.C. Quillen, M. Williams, et al., astroph:
0812.4032 (2009).

41. J.F. Navarro, A. Helmi, and K. Freeman, Astrophys. J. 601, L43
(2004).

42. B. Nordstr\"{o}m, M. Mayor, J. Andersen, et al., Astron.
Astrophys. 419, 989 (2004).

43. R.P. Olling and W. Dehnen, Astrophys. J. 599, 275 (2003).

44. A.C. Quillen and I. Minchev, Astron. J. 130, 576 (2005).

45. W. Schuster and P. Nissen, Astron. Astrophys. 221, 65 (1989).

46. W. Schuster, A.Moitinho, A.Marquez, et al., Astron. Astrophys.
445, 939 (2006).

47. G.M. De Silva, K.C. Freeman, J. Bland-Hawthorn, et al.,
Astron. J. 133, 694 (2007).

48. B.W. Silverman, Density Estimation for Statistics and Data
Analysis (Chapman Hall, London, 1986).

49. R.S. De Simone, X. Wu, and S. Tremain, Mon. Not. R. Astron.
Soc. 350, 627 (2004).

50. J. Skuljan, J.B. Hearnshaw, and P.L. Cottrell, Mon. Not. R.
Astron. Soc. 308, 731 (1999).

51. D.R. Soderblom, B.F. Jones, S. Balachandran, et al., Astron.
J. 106, 1059 (1993).

52. C. Soubiran, O. Bienaym\'e , T.M. Mishenina, et al., Astron.
Astrophys. 480, 91 (2008).

53. M. Steinmetz, T. Zwitter, A. Seibert, et al., Astron. J. 132,
1645 (2006).

54. B.J. Taylor, Astron. Astrophys. 362, 563(2000).

55. J. Torra, D. Fern\'andez, and F. Figueras, Astron. Astrophys.
359, 82 (2000).

56. V.V. Vityazev, Wavelet Analysis of Time Series (SPb. Gos.
Univ., St.-Petersbourg, 2001) [in Russian].

57. S. K. Yi, Y.-C. Kim, and P. Demarque, Astrophys. J. Suppl.
Ser. 144, 259 (2003).

58. M.V. Zabolotskikh, A.S. Rastorguev, and A.K. Dambis, Pis'ma
Astron. Zh. 28, 516 (2002) [Astron. Lett. 28, 454 (2002)]

\newpage
{
\begin{table}[t]                                                
\caption[]{\small\baselineskip=1.0ex\protect
 Characteristics of the Wolf 630-$\alpha$Ceti branch
(features W1 and W2) and the Hercules stream (features H1 and H2)

}
\begin{center}
\begin{tabular}{|r|r|c|c|c|r|r|r|r|r|r|}\hline

 Obj. & $N_\star$ &     [Fe/H],    &   Age,      & $ U,$  & $ V,$ & $ W,$ & $ a_i,$ & $ b_i,$ & $\beta_i$ \\
      &           &        dex     &   Gyr       &  km s$^{-1}$  & km s$^{-1}$  &  km s$^{-1}$ &  km s$^{-1}$  & km s$^{-1}$  & deg.  \\\hline
   W1 &     88    & $-0.06~(0.20)$ & $3.9~(2.7)$ & $  23$ & $-28$ & $ -5$ & $ 7.4$ & $5.6$ & $148^\circ$ \\\hline
   W2 &     95    & $-0.13~(0.19)$ & $3.6~(2.3)$ & $  41$ & $-26$ & $ -8$ & $ 8.9$ & $6.3$ & $120^\circ$ \\\hline
   H1 &    136    & $-0.09~(0.17)$ & $4.6~(3.2)$ & $-33~$ & $-51$ & $ -8$ & $14.2$ & $5.4$ & $103^\circ$ \\\hline
   H2 &     71    & $-0.16~(0.27)$ & $5.7~(3.4)$ & $-77~$ & $-49$ & $ -7$ & $21.2$ & $7.9$ & $ 80^\circ$ \\\hline
\end{tabular}
\end{center}
 {\small
Note. $N$ is the number of stars with available age and
metallicity estimates, the velocities $U,V,$ and $W$ are given
relative to the Sun (see Fig. 4), the corresponding dispersions
are given for the mean metallicity indices and mean ages of the
sample stars.
 }
 \vskip 80mm
\end{table}
}
{
\begin{table}[t]                                                
\caption[]{\small\baselineskip=1.0ex\protect
 Parameters of the Hipparcos stars that are probable
members of the KFR08 stream

}
\begin{center}
\begin{tabular}{|r|c|c|c|c|r|r|r|c|l|l|}\hline

    HIP &   & [Fe/H]  & Ref & Age & $U\pm e_U$  & $V\pm e_V$  & $W\pm e_W$  & $p$  \\\hline
   5336 &   & $-0.84$ & (1) &     & $ -32\pm1~$ & $-153\pm1~$ & $ -28\pm1~$ & 1.00 \\\hline
  15495 &   & $-0.36$ & (2) &     & $  58\pm4~$ & $-174\pm8~$ & $  -3\pm3~$ & 1.00 \\\hline
  18235 &   & $-0.71$ & (3) & 11  & $ -16\pm3~$ & $-161\pm4~$ & $ -19\pm2~$ & 1.00 \\\hline
  19143 &   & $-0.49$ & (2) &     & $-140\pm3~$ & $-143\pm11$ & $ -42\pm2~$ & 0.98 \\\hline
  54469 & * & $-0.72$ & (4) & 11  & $  91\pm5~$ & $-159\pm16$ & $ -64\pm16$ & 1.00 \\\hline
  55988 &   &         &     &     & $  50\pm4~$ & $-154\pm6~$ & $ -25\pm4~$ & 0.99 \\\hline
  58357 & * & $-0.71$ & (1) &     & $-123\pm16$ & $-134\pm23$ & $  45\pm1~$ & 0.65 \\\hline
  58708 &   & $-0.30$ &     &     & $ -14\pm3~$ & $-160\pm4~$ & $  15\pm1~$ & 0.99 \\\hline
  58843 &   & $-0.80$ &     &     & $ 122\pm9~$ & $-138\pm14$ & $ -58\pm7~$ & 0.81 \\\hline
  59785 &   & $-0.37$ &     &     & $-117\pm9~$ & $-136\pm6~$ & $-109\pm7~$ & 0.92 \\\hline
  60747 & * & $-0.77$ & (6) &     & $ 110\pm7~$ & $-146\pm14$ & $  91\pm7~$ & 0.91 \\\hline
  64920 &   & $-0.42$ &     &     & $  66\pm5~$ & $-159\pm5~$ & $  43\pm5~$ & 0.99 \\\hline
  74033 &   & $-0.75$ & (4) & 13  & $-113\pm10$ & $-132\pm10$ & $  42\pm7~$ & 0.65 \\\hline
  81170 & * & $-1.26$ & (5) &     & $ -77\pm2~$ & $-157\pm9~$ & $-123\pm3~$ & 0.99 \\\hline
  87101 &   & $-1.31$ & (6) &     & $ -76\pm5~$ & $-159\pm18$ & $  -3\pm2~$ & 0.91 \\\hline
  93269 &   &         &     &     & $  70\pm3~$ & $-140\pm1~$ & $  -4\pm3~$ & 0.99 \\\hline
  93623 &   & $-0.60$ & (2) &     & $ 130\pm5~$ & $-149\pm16$ & $ -20\pm1~$ & 0.96 \\\hline
  96185 &   & $-0.60$ & (4) & 12  & $ -56\pm1~$ & $-156\pm1~$ & $  66\pm1~$ & 1.00 \\\hline
 117702 &   & $-0.43$ & (7) &     & $  12\pm7~$ & $-159\pm7~$ & $ 124\pm5~$ & 0.99 \\\hline

\end{tabular}
\end{center}
 {\small
Note. The age is in Gyr, the velocities $U,V,$ and $W$ are in km
s$^{-1}$ and are given relative to the LSR (Dehnen and Binney
1998); the asterisk $*$ marks the candidates with
$e_\pi/\pi<0.15;$ the stellar metallicities and age estimates were
taken from the following papers: 1, Soubiran et al. (2008); 2,
Ibukiyama, and Arimoto (2002); 3, Bensby et al. (2005); 4,
Holmberg et al. (2007); 5, Borkova and Marsakov (2005); 6,
Schuster et al. (2006); 7, Jenkins et al. (2008).
 }
\end{table}
}

\newpage
\begin{figure}[p]
{
\begin{center}
 \includegraphics[width=100mm]{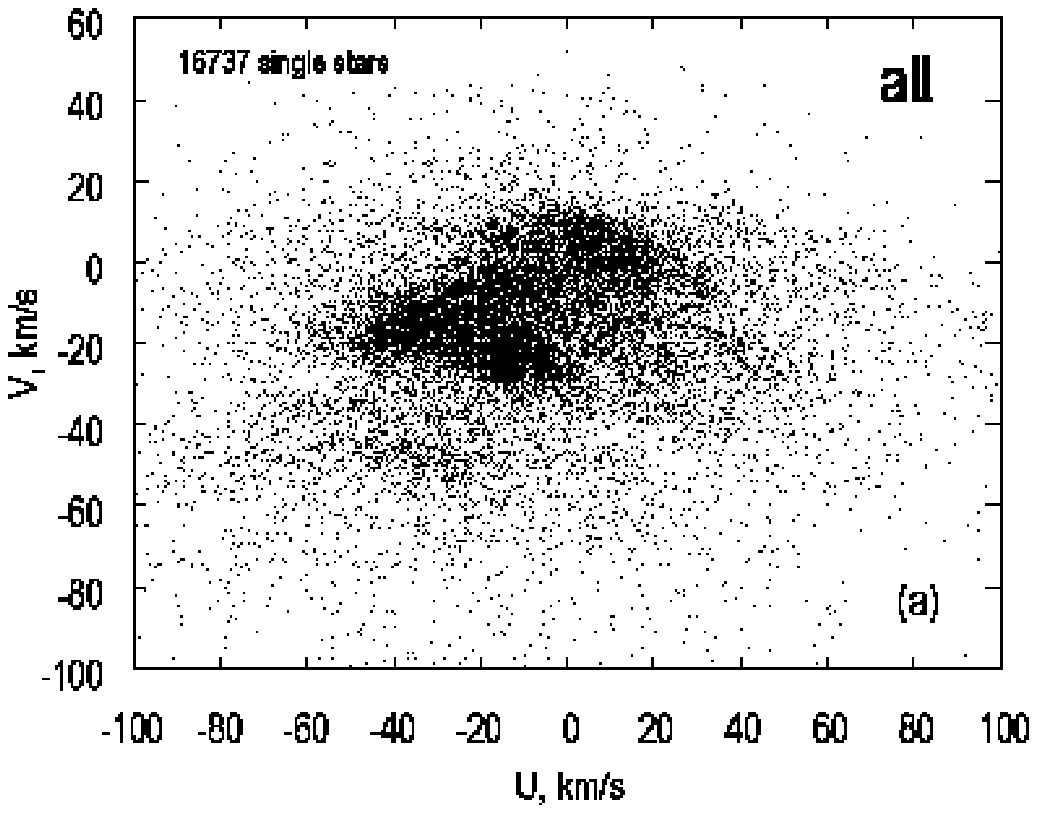}

 \includegraphics[width=95mm]{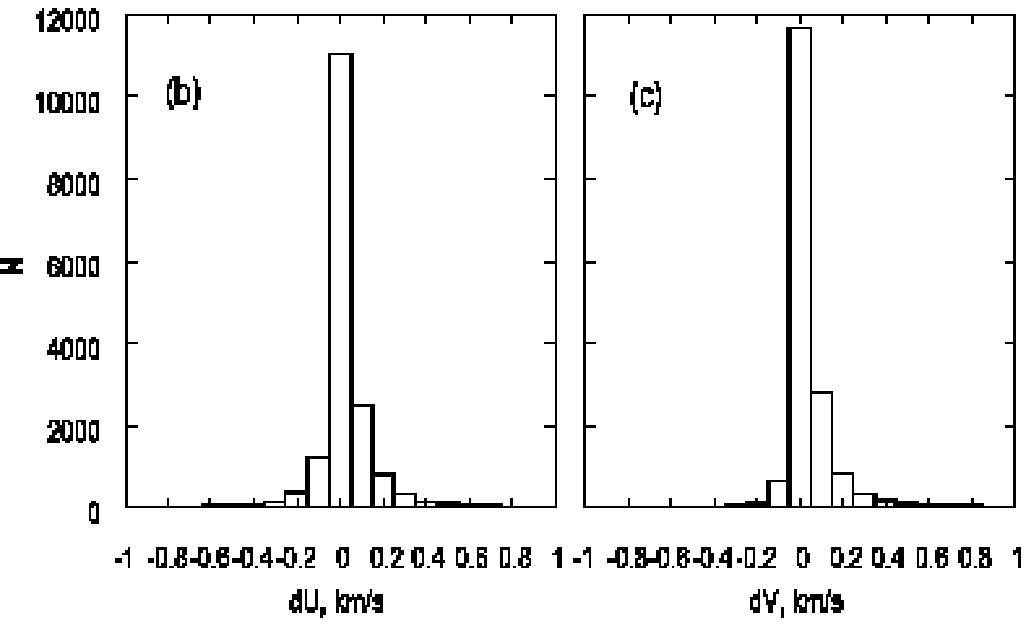}
\end{center}
} Fig.~1. (a) $UV$ velocity distribution for the ``all'' sample of
16737 single stars with reliable distance estimates
$(e_\pi/\pi<0.1);$ the velocities are given relative to the Sun.
Distributions of the (b) $U$ and (c) $V$ velocity biases caused by
the measurement errors of the stellar parallaxes.
\end{figure}

\newpage
\begin{figure}[p]
{
\begin{center}
 \includegraphics[width=100mm]{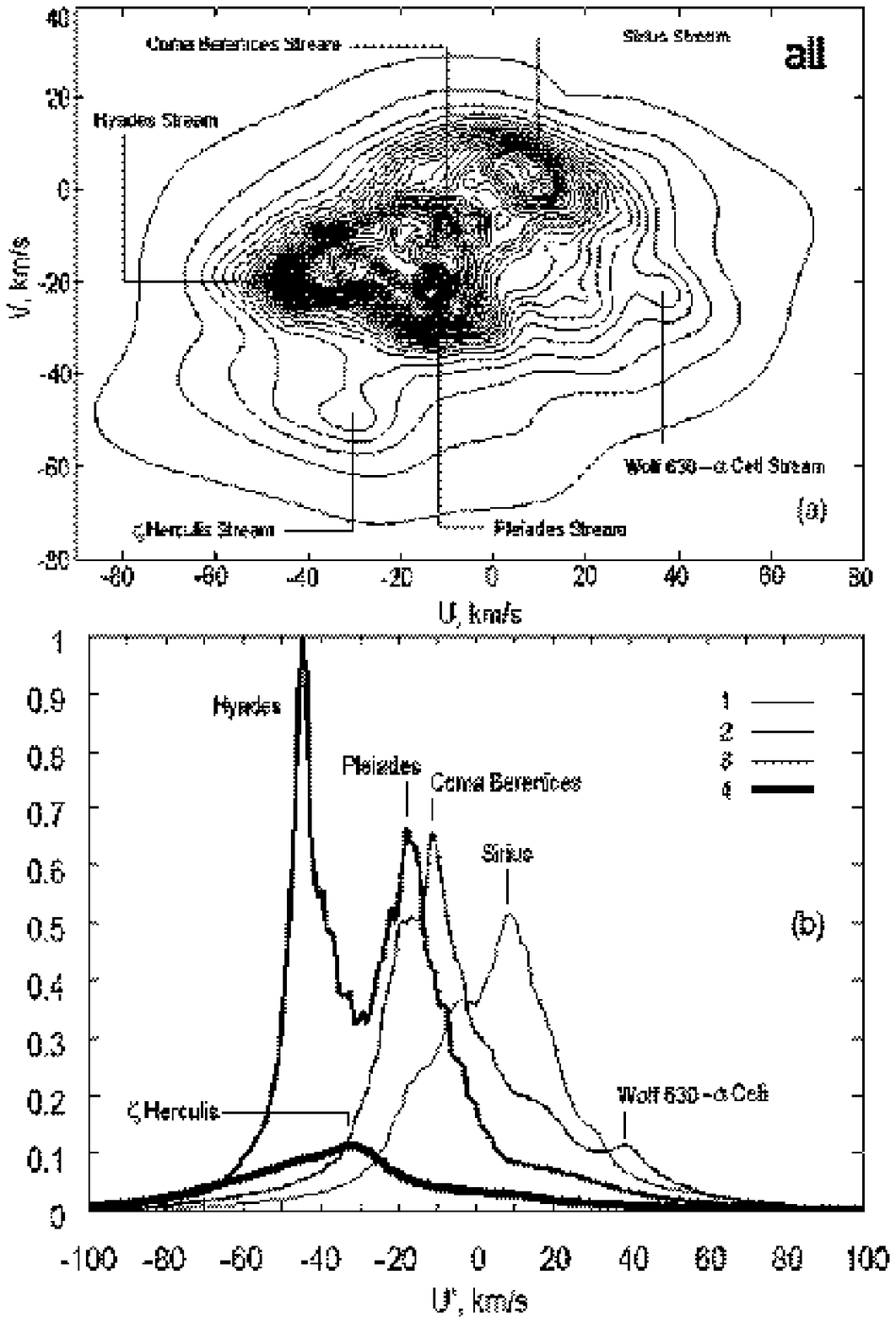}
\end{center}
} Fig.~2. Density of the $UV$-velocity distribution corresponding
to Fig. 1 obtained by the adaptive kernel method; the velocities
are given relative to the Sun (a); the sections of map (a)
perpendicular to the $(U,V)$ plane that pass through the main
peaks and that make $+16^\circ$ with the $U$ axis if measured
clockwise (this axis is designated as $U);$ the distribution
density in units of $7 \times 10^{-4}$ is along the vertical axis,
the numbers denote the sections passing through the Sirius (1),
Coma Berenices (2), Pleiades-Hyades (3), and Hercules (4)
branches)(b).
 \end{figure}

\newpage
\begin{figure}[p]
{
\begin{center}
 \includegraphics[width=100mm]{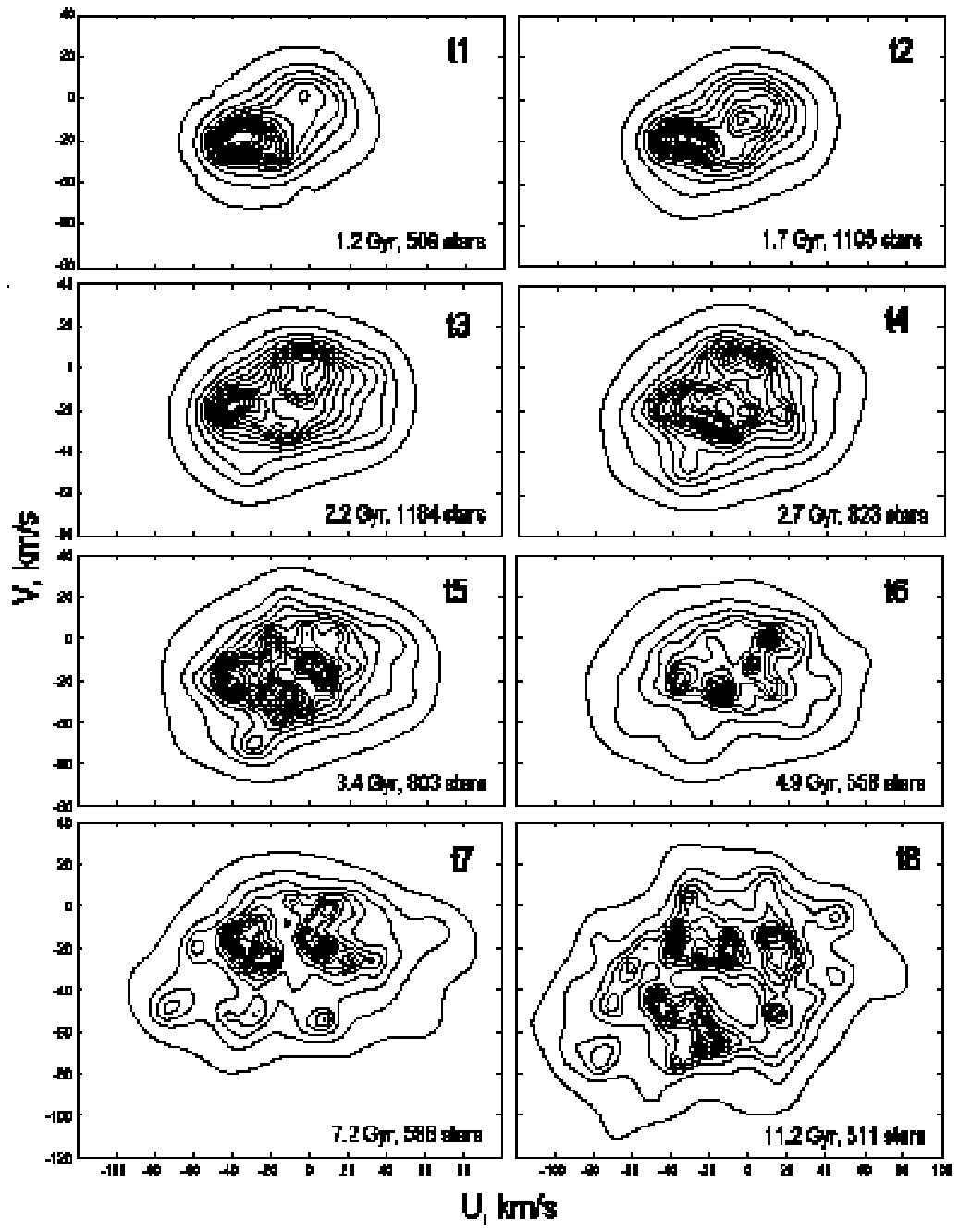}
\end{center}
} Fig.~3. Densities of the $UV$-velocity distribution for samples
of F and G dwarfs as a function of the stellar age; the velocities
are given relative to the Sun.
\end{figure}

\newpage
\begin{figure}[p]
{
\begin{center}
 \includegraphics[width=100mm]{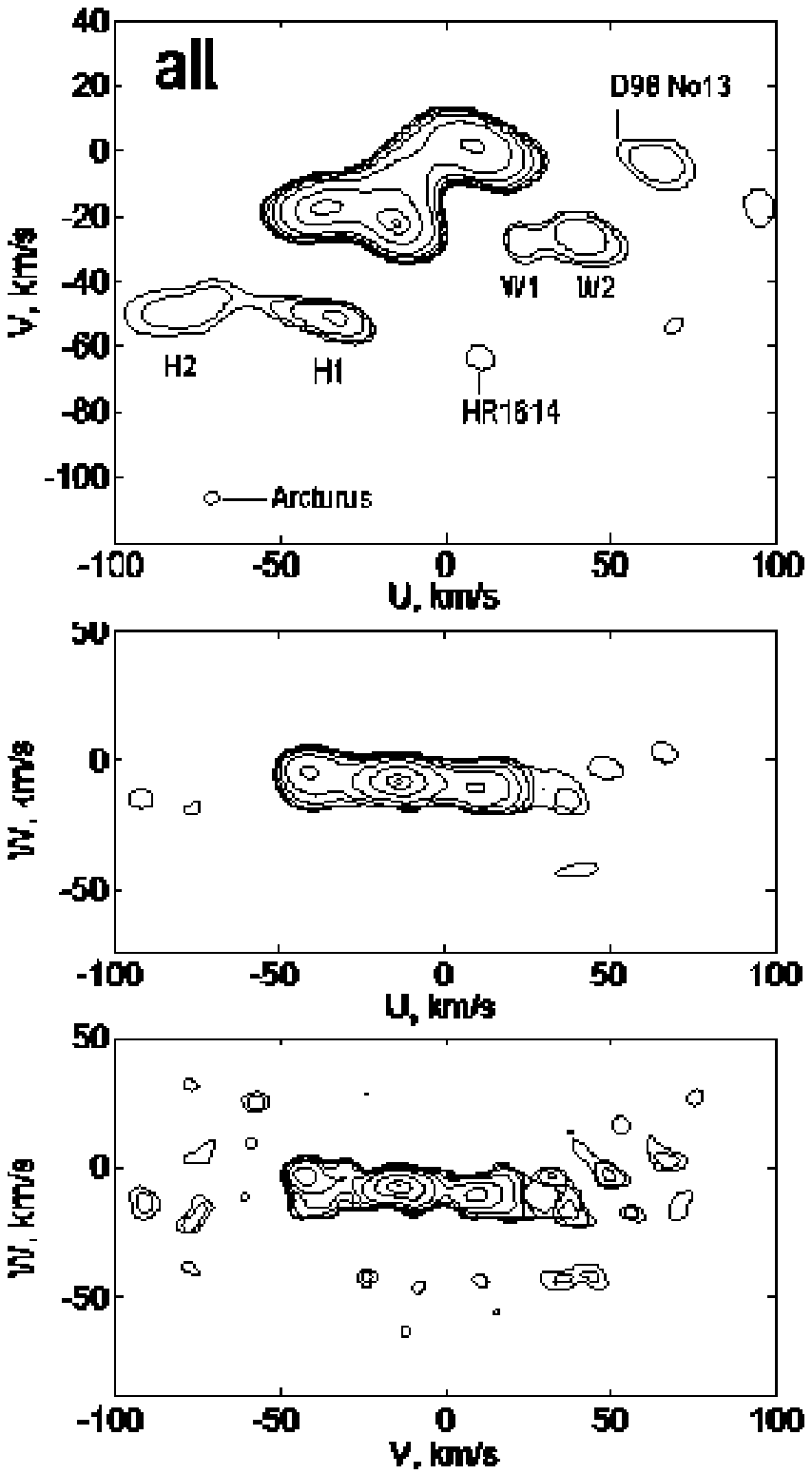}
\end{center}
} Fig.~4. Wavelet maps of $UV,WU,$ and $WV$ velocities for a
sample of 16737 stars; the velocities are given relative to the
Sun. See also the text.
\end{figure}

\newpage
\begin{figure}[p]
{
\begin{center}
 \includegraphics[width=110mm]{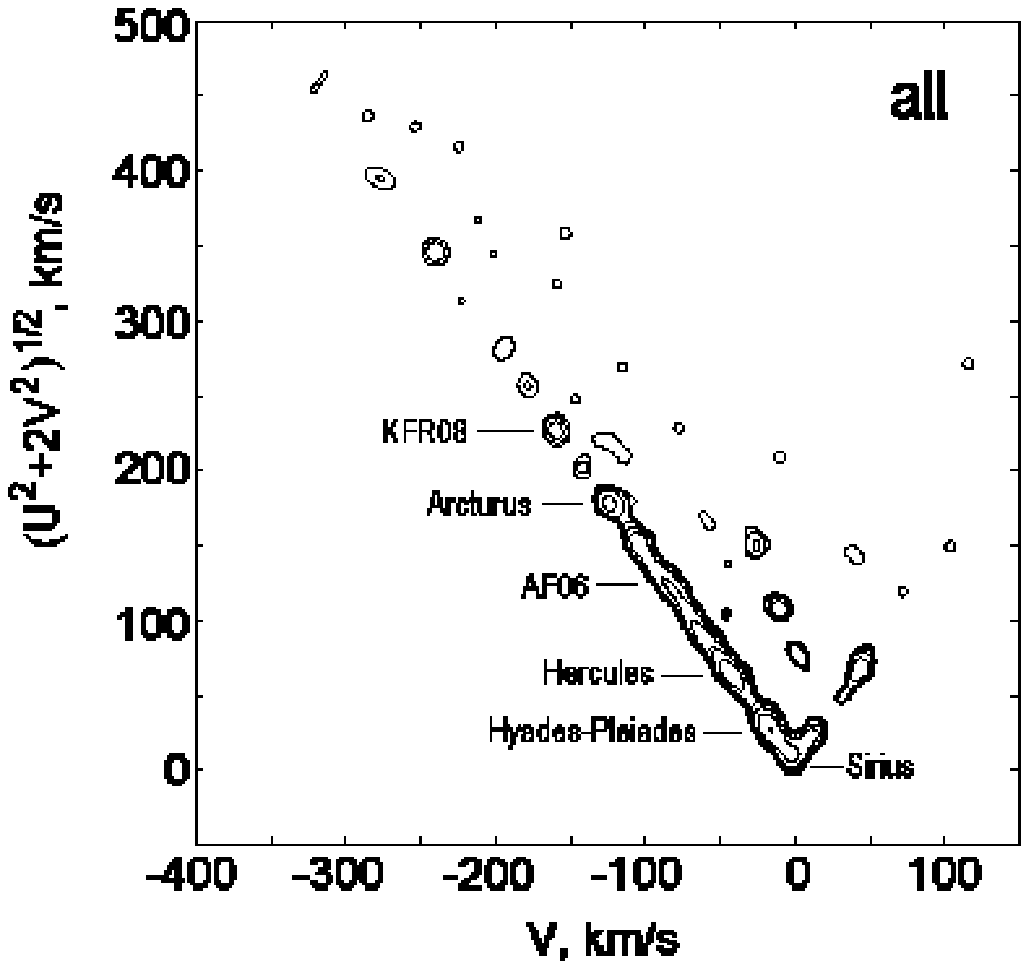}
\end{center}
} Fig.~5. Wavelet maps in the system of $(V, \sqrt {U^2 + 2V^2} )$
coordinates for a sample of 16737 stars; the velocities are given
relative to the LSR.
\end{figure}

\newpage
\begin{figure}[p]
{
\begin{center}
 \includegraphics[width=110mm]{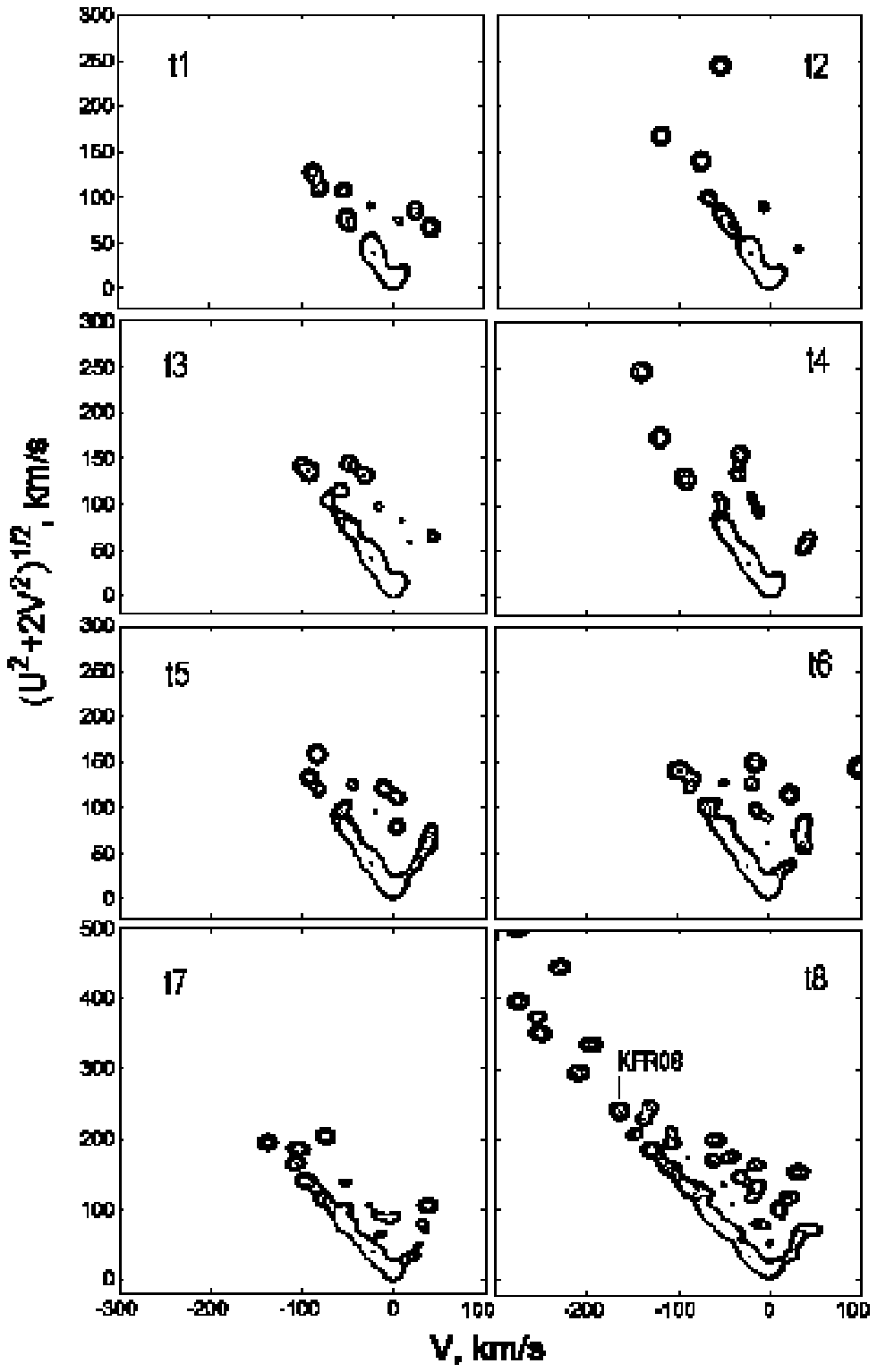}
\end{center}
} Fig.~6. Wavelet maps in the system of $(V, \sqrt {U^2 + 2V^2} )$
coordinates for samples of F and G dwarfs as a function of the
stellar age; the velocities are given relative to the LSR.
\end{figure}

\newpage
\begin{figure}[p]
{
\begin{center}
 \includegraphics[width=100mm]{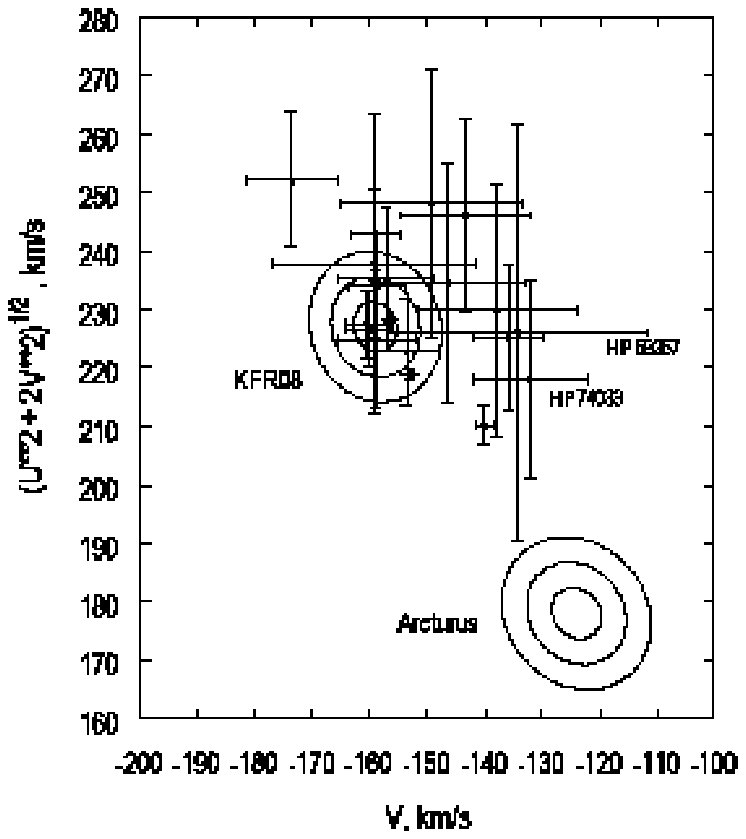}
\end{center}
} Fig.~7. Positions of KFR08 stream members in the
 $(V,\sqrt{U^2+2V^2})$ plane, the velocities are given relative to the LSR,
three contours corresponding to probabilities of 0.683, 0.954, and
0.997 $(1\sigma,2\sigma,3\sigma)$ are given for the KFR08 and
Arcturus streams.
\end{figure}

\newpage
\begin{figure}[p]
{
\begin{center}
 \includegraphics[width=100mm]{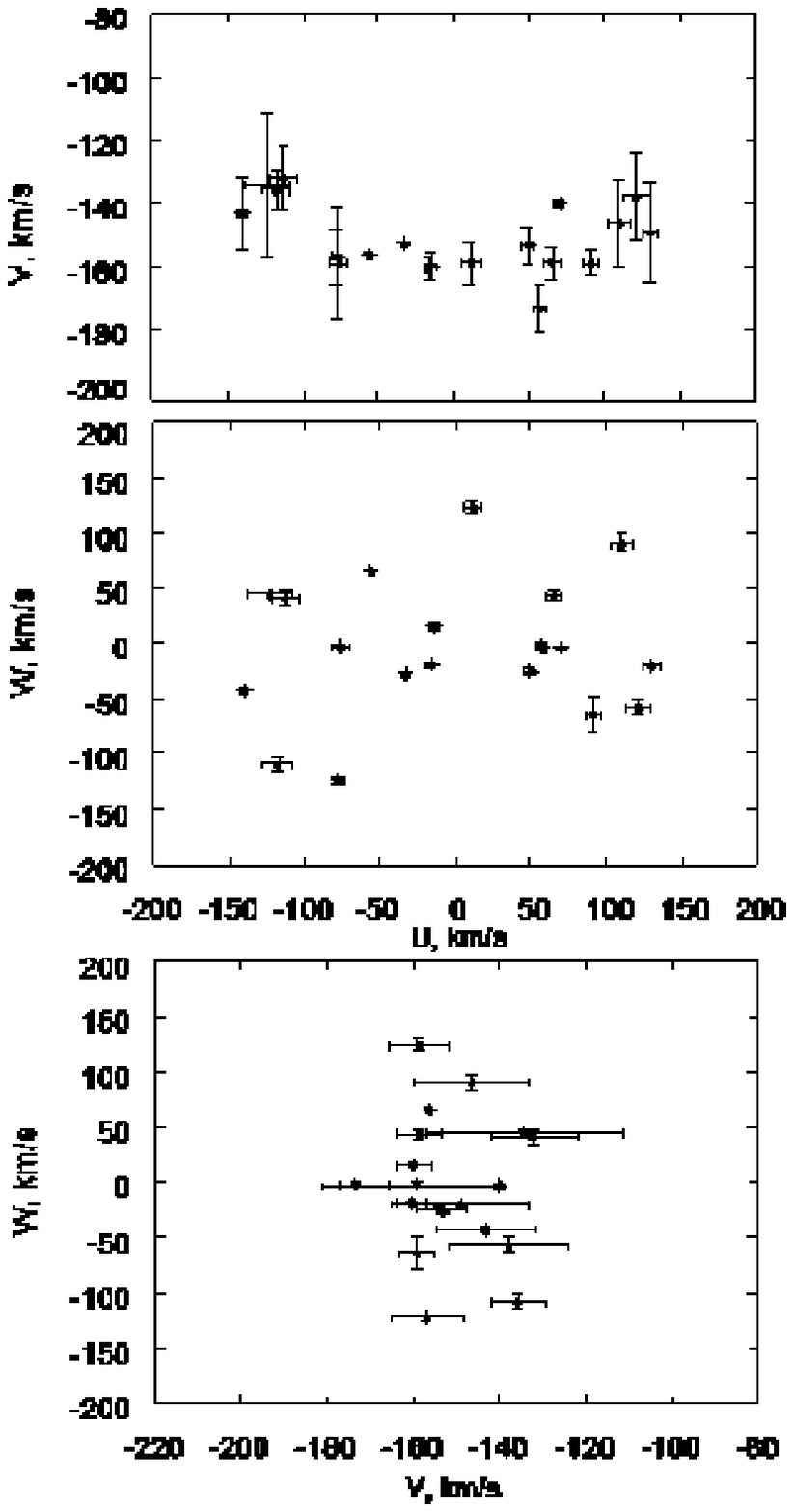}
\end{center}
} Fig.~8. Velocity distribution for KFR08 stream members, the
velocities are given relative to the LSR.
\end{figure}

\newpage
\begin{figure}[p]
{
\begin{center}
 \includegraphics[width=100mm]{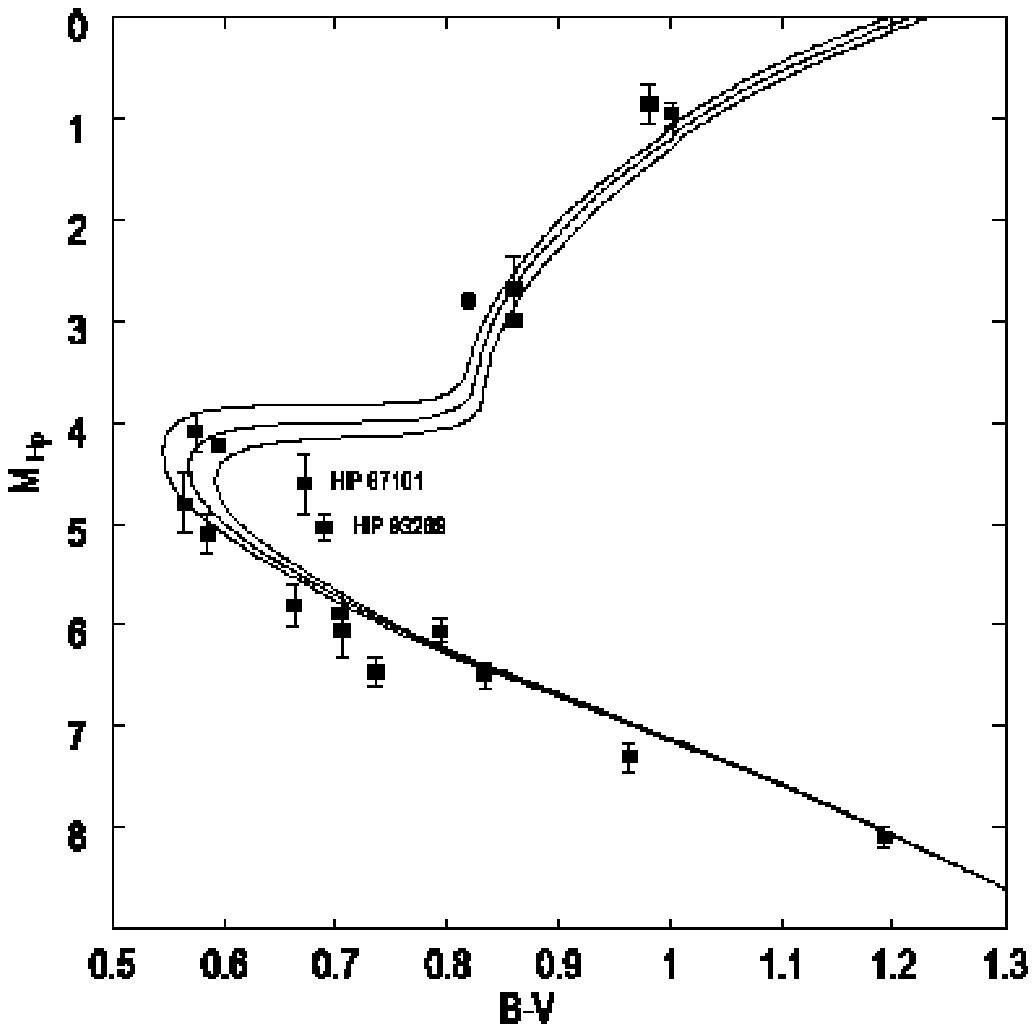}
\end{center}
} Fig.~9. Color-absolute magnitude diagram for KFR08 stream
members.
\end{figure}

\end{document}